\documentclass[%
 reprint,
 superscriptaddress,
 amsmath,
 aps, prl,
]{revtex4-1}

\usepackage{graphicx}
\usepackage{dcolumn}
\usepackage{bm}

\begin{document}

\preprint{APS/123-QED}

\title{Topological spin excitations in honeycomb ferromagnet CrI$_3$}

\author{Lebing Chen}
\affiliation{Department of Physics and Astronomy,
Rice University, Houston, Texas 77005, USA}
\author{Jae-Ho Chung}
 \email{jaehc@korea.ac.kr}
\affiliation{Department of Physics, Korea University, Seoul 02841, Korea}%
\affiliation{Department of Physics and Astronomy,
Rice University, Houston, Texas 77005, USA}
\author{Bin Gao}
\affiliation{Department of Physics and Astronomy,
Rice University, Houston, Texas 77005, USA}
\author{Tong Chen}
\affiliation{Department of Physics and Astronomy,
Rice University, Houston, Texas 77005, USA}
\author{Matthew B. Stone}
\author{Alexander I. Kolesnikov}
\affiliation{Neutron Scattering Division, Oak Ridge National Laboratory, Oak Ridge, Tennessee 37831, USA}
\author{Qingzhen Huang}
\affiliation{NIST Center for Neutron Research, National Institute of Standards and Technology, Gaithersburg, MD 20899-6102, USA}
\author{Pengcheng Dai}
 \email{pdai@rice.edu}
\affiliation{Department of Physics and Astronomy,
Rice University, Houston, Texas 77005, USA}

\date{\today}

\begin{abstract}
In two dimensional honeycomb ferromagnets, bosonic magnon quasiparticles (spin waves) may either behave as massless Dirac fermions or form topologically protected edge states. The key ingredient defining their nature is the next-nearest neighbor Dzyaloshinskii-Moriya (DM) interaction that breaks the inversion symmetry of the lattice and discriminates chirality of the associated spin-wave excitations. Using inelastic neutron scattering, we find that spin waves of the insulating honeycomb ferromagnet CrI$_3$ ($T_C=61$ K) have two distinctive bands of ferromagnetic excitations 
separated by a $\sim$4 meV gap at the Dirac points.
These results can only be understood by considering a Heisenberg Hamiltonian with DM interaction, thus providing experimental evidence that spin waves in CrI$_3$ can have robust topological properties potentially useful for dissipationless spintronic applications.
\end{abstract}

\maketitle


\section{Introduction}

When quantum particles such as electrons are confined in two-dimensional (2D) geometry, 
the reduced lattice dimensions and particle interactions can drive the system into novel behavior 
such as the quantum Hall state under a large magnetic
field perpendicular to the 2D electron gas \cite{Klitzing}. In 1988, Haldane showed that some solid state systems, for example the 2D honeycomb lattice, can also have quantum Hall state without having to apply magnetic fields due to their inherently topological band structure \cite{Haldane}. 
It is well known that materials with strong spin-orbit coupling can 
host topological band structures \cite{Hasan_2010,Qi}.  
For 2D honeycomb and kagome lattices, a diverse range of novel 
electronic band properties and magnetism
have been observed \cite{Bansil,yizhou}. For instance, a graphene as the simplest honeycomb exhibits linear electronic dispersions near the Fermi surface allowing exotic massless Dirac fermions to appear \cite{CastroNeto_2009,Wehling_Dirac_2014}. Such band structure is built upon the two equivalent and interconnected triangular sublattices, which result in topological band crossing at the Fermi surface [See Figs. 1(a) and 1(b) for the real and reciprocal spaces, respectively, of the honeycomb lattice.].  

Topological band structures are not unique to systems
with odd half-integer spin electron-like quasiparticles (fermions) like graphene.  In fact, many systems with integer spin quasiparticles (bosons) can also have topological
band structures. For example, topological photon modes have been realized in photonic crystals \cite{Raghu,ZWang2009}. In addition, 
anomalous thermal Hall effect from topological magnon band structures have been
predicted in insulating quantum magnets \cite{Katsura_2010} and observed 
in an insulating collinear ferromagnet Lu$_2$V$_2$O$_7$ with a pyrochlore structure \cite{Onose_2010}. 
Theoretically, several classes of ferromagnetic insulators have been predicted 
to have interesting topological properties \cite{Zhang_2013,Shindou_2013,Mook2014,Mook2016,FYLi}. In the case of 2D honeycomb ferromagnets with two magnetic atoms per unit cell,
 magnetic versions of Dirac particles have been predicted \cite{Fransson_Dirac_2016,Owerre_CrX3_2017,Pershoguba_CrBr3_2018}. The magnon (spin-wave) band structure of these ferromagnets are essentially identical to the electronic counterpart of graphene with two modes, acoustic and optical spin waves, for each state reflecting two sublattices. If the spins interact only via the Heisenberg exchange couplings, the two spin wave modes will cross with each other at $K/K^\prime$ points at the corner of the Brillouin Zone (BZ) boundary and form Dirac cones with linear dispersion \cite{Fransson_Dirac_2016,Owerre_CrX3_2017,Pershoguba_CrBr3_2018}. The presence of these Dirac points are robust against finite next-nearest neighbor exchanges, which will only shift positions of the Dirac points. Such spin wave bands have experimentally been observed in 2D ferromagnets CrBr$_3$ and Cr$_2$Si$_2$Te$_6$ \cite{Samuelsen_CrBr3_1971,Williams_CrSiTe3_2015}, thus confirming the presence of non-degenerate band-touching (Dirac) points in the magnon excitation spectrum and leading to a massless Dirac Hamiltonian \cite{Fransson_Dirac_2016,Owerre_CrX3_2017,Pershoguba_CrBr3_2018}. Similar magnon band crossings have also been observed in the three-dimensional (3D) antiferromagnet Cu$_3$TeO$_6$ \cite{Yao_Cu3TeO6_2017,Bao_Cu3TeO6_2018}. 

\begin{figure}[b]
\centering
\includegraphics[width=3.33in]{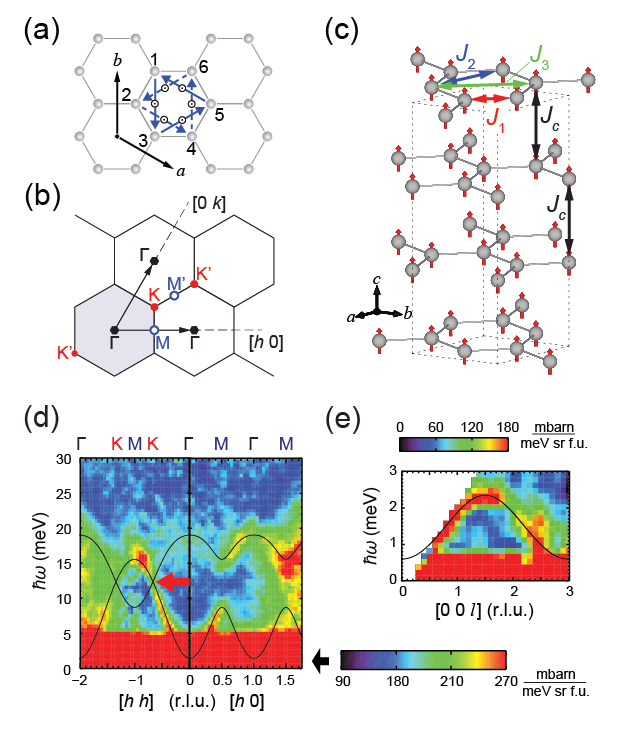}  
\caption{\label{fig1} (a) Top view of the honeycomb lattice and the out-of-plane components of the antisymmetric DM vectors. The triangular arrows mark the second nearest neighbor bond orientations that share the common sign of DM vectors. (b) The BZ zone boundaries and the high symmetry points on the $(h,k)$ plane. (c) Crystal and magnetic structures of CrI$_3$ with
the Heisenberg exchange paths. Iodine sites are not displayed for simplicity. The dotted lines denote the hexagonal unit cell boundary. (d) Inelastic neutron scattering results of spin waves with $E_i = 50$ meV along the high symmetry directions in the $(h,k)$ plane, which are observed at 5 K and integrated over $-5\leq l \leq 5$. The superimposed solid lines are the calculations by the Heisenberg-only model. (e) Low-energy spin-wave mode along the $l$ direction ($E_i = 8$ meV). Note that the minima appear at $l = 3n$ in the hexagonal setting. The data are normalized to the absolute units
of mbarn/meV/Sr/f.u. by a vanadium standard.}
\end{figure}

In the case of graphene, a finite spin-orbit coupling produces a small bulk semiconducting gap ($\sim$1 $\mu$eV), leaving only the edge states to be truly conducting at absolute zero temperature \cite{Yao_graphene_2007,Hasan_2010}. A magnetic version of such topological edge states may also be realized if strong spin-orbit coupled antisymmetric Dzyaloshinkii-Moriya (DM) exchange opens a gap at the spin-wave crossing Dirac points \cite{Owerre_topological_2016,SKKim_HKM_2016,Ruckriegel_2018}. The DM interactions are known to act as effective vector potential leading to anomalous magnon Hall effect that facilitates topological edge transports \cite{Katsura_2010,Onose_2010,Cheng_2016}.
In contrast to electron spin current where dissipation can be large due to Ohmic heating, noninteracting topological magnons, which are quantized spin-1 excitations from an ordered magnetic ground state, are uncharged and can in principle propagate for a long time without dissipation \cite{Chumak,Chernyshev2016,XSWang2017,XSWang2018}.
Since the DM interaction will cancel out upon space inversion, a finite DM term may appear only between the next-nearest neighbors on the 
honeycomb lattices [see Fig. \ref{fig1}(a)]. Whereas the possible orientations of these DM vectors may depend on local symmetries \cite{Moriya_1960}, only the term collinear with magnetic moments can induce the Dirac gap. Such DM-induced topological magnons have been predicted \cite{Zhang_2013,Pereiro_2014} and 
observed in 2D kagome ferromagnet compound Cu[1,3-benzenedicarboxylate(bdc)] (Cu(1,3-bdc)), where an out-of-plane external magnetic field applied 
to align the in-plane ferromagnetic ordered moments
along the $c$-axis is found to also induce a spin gap
at the Dirac points \cite{Chisnell_kagome_2015,Chisnell2016}.

In honeycomb ferromagnets, it is unclear whether such topological magnons can exist. In fact, the topology of the next-nearest neighbor bonds on a honeycomb lattice is equivalent to the nearest neighbor bonds of a kagome lattice. Recent experimental discoveries of intrinsic 2D ferromagnetism in van der Waals materials suggest that the topological spin excitations will probably be more robust in the honeycomb lattices \cite{Gong_Ge2Cr2Te6_2017,Huang_CrI3_2017,Bonilla_VSe2_2018}


In this work, we use inelastic neutron scattering to map out energy and wave vector dependence of 
spin-wave excitations in CrI$_3$, one of the honeycomb ferromagnets where topological Dirac magnons are predicted to appear \cite{Pershoguba_CrBr3_2018,Owerre_topological_2016}. The  honeycomb lattice, shown in Fig. \ref{fig1}(c), is essentially identical to those of another chromium trihalide CrBr$_3$, in which spin-wave excitations have long been known \cite{Samuelsen_CrBr3_1971}. The magnetism in CrI$_3$ is commonly ascribed to Cr$^{3+}$ ions surrounded by I$_6$ octahedra, forming a 2D honeycomb network [Fig. \ref{fig1}(c)]. The CrI$_3$ layers are stacked against each other by van der Waals interaction, and have a monoclinic crystal structure at room temperature. Upon cooling, the monoclinic crystal structure transforms to the rhombohedral structure (space group: $R\bar{3}$) over a wide temperature range (100 - 220 K) via lateral sliding of the CrI$_3$ planes with hysteresis \cite{McGuire_CrI3_2014}. At Curie temperature $T_C = 61$ K, ferromagnetic ordering appears with Cr$^{3+}$ spins oriented along the $c$ axis [Fig. 1(c)] \cite{SI}. This is different from the in-plane moment
of Cu(1,3-bdc) at zero field, providing the necessary condition for DM interactions to open a gap in CrI$_3$ without the need for an external magnetic field \cite{Zhang_2013,Pereiro_2014}. 
Since CrI$_3$ has similar structural and ferromagnetic transitions as that of CrBr$_3$
 albeit at different temperatures, one would expect that 
spin-wave excitations of CrI$_3$ should be similar to that of CrBr$_3$, which have Dirac points at the acoustic and optical spin-wave crossing points \cite{Samuelsen_CrBr3_1971}. Surprisingly, we find that spin waves in CrI$_3$ exhibit remarkably large gaps at the Dirac points, thus providing direct evidence for the presence 
of DM interactions in CrI$_3$ \cite{Owerre_topological_2016}. Therefore, CrI$_3$ is an insulating ferromagnet that can potentially host topological edge magnons predicted by the theories \cite{Owerre_topological_2016,SKKim_HKM_2016,Ruckriegel_2018}. 

\section{Results and Discussions}

Thin single crystal platelets of CrI$_3$ were grown by the chemical vapor transport method using I$_2$ as the transport agent \cite{McGuire_CrI3_2014}. The grown crystals are typically 1 cm by 1 cm in area and extremely thin and fragile \cite{SI}. 
Our results are reported using a honeycomb structure with 
in-plane Cr-Cr distance of $\sim$3.96 \AA\ and $c$-axis layer spacing of 6.62 \AA\ in the low temperature rhombohedral structure [Fig. \ref{fig1}(c)] \cite{MAM2017}. The in-plane momentum transfer $\textbf{q}=h\textbf{a}^\ast+k\textbf{b}^\ast$ is denoted as $\textbf{q}=(h,k)$ in hexagonal reciprocal lattice units (r.l.u.) as shown in Fig. 1(b). 
The temperature-dependent magnetization and neutron powder diffraction measurements confirmed that the ferromagnetic transition occurs at $T_C \approx 61$ K with an ordered moment of $3.0\pm 0.2$ $\mu_B$ per Cr$^{3+}$ at 4 K, and the magnetic anisotropy has an easy-axis along the $c$ axis \cite{SI}. To observe spin-wave excitations, we co-aligned and stacked $\sim$25 pieces of platelets with a total mass of $\sim$0.3-g. Time-of-flight inelastic neutron scattering experiments were performed using the SEQUOIA spectrometer of the Spallation Neutron Source at the Oak Ridge National Laboratory using three different incident energies of 
$E_i = 50, 25,$ and 8 meV \cite{Sequoia}. Neutron powder diffraction measurements were carried out using the BT-1 diffractometer of NIST Center for Neutron Research. 

Figure \ref{fig1}(d) shows an overview of spin-wave dispersions along high symmetry directions in the $(h,k)$ plane. A nearly isotropic spin-wave mode emerges from the $\Gamma$ point at the ferromagnetic zone center and moves towards the zone boundary with increasing energy. This low-energy mode accounts for the in-phase oscillations between the two sublattice Cr spins within a unit cell. In the rest of the manuscript, we refer to this low energy mode as the ``acoustic'' magnon mode. Along the $[h,0,0]$ direction towards the $M$ point [Fig. 1(b)], the acoustic mode reaches its maximum energy around $\hbar\omega=10$ meV while another mode is visible at high energy between 15 meV and 19 meV. 
This high energy mode accounts for the out-of-phase oscillations between the two sublattice Cr spins, which we refer to as the ``optical'' mode. The large separation in energy between the two modes is consistent with the dominant ferromagnetic exchanges.
Using a simple Heisenberg Hamiltonian with only
the in-plane magnetic exchange couplings and without the DM interaction \cite{Owerre_topological_2016}, we can fit to the overall momentum dependence of the spin-wave excitations as 
the solid lines in Fig. \ref{fig1}(d) \cite{SI}. 
While the overall agreement of the Heisenberg Hamiltonian is reasonably good, the calculation apparently fails to explain the observed spin-wave dispersions along the $[h,h,0]$ direction going through the Dirac $K$ point. 
As indicated by a thick red arrow in Fig. \ref{fig1}(d), the spin-wave intensity exhibits a clear discontinuity where the acoustic and optical modes are expected to cross each other. This observation strongly suggests that magnons at the Dirac points in CrI$_3$ have a finite effective mass contrary to the Heisenberg-only Hamiltonian \cite{Owerre_topological_2016}.

\begin{figure}[b]
\centering
\includegraphics[width=3.33in]{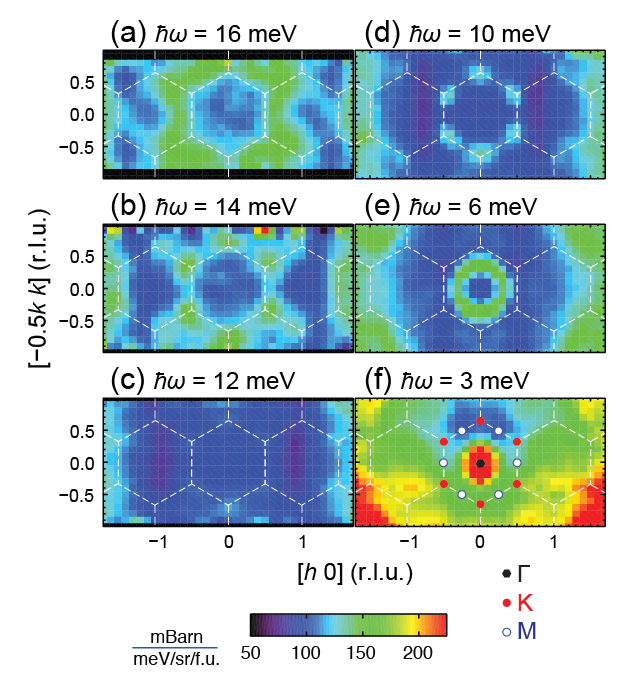}
\caption{\label{fig2} Constant-$\hbar\omega$ cuts of spin waves at selected energy transfer values within the $(h,k)$ plane at 5 K. In all plots, data were subtracted by the empty can background, integrated over the energy range of $\hbar\omega\pm 0.5$ meV as well as integrating over $-5 \le l \le 5$. 
The dashed lines are the BZ boundaries of the 2D reciprocal lattice. In (f), $\Gamma$, $K$ and $M$ points are marked along the 1st BZ boundary for clarity.}
\end{figure}

To accurately determine the spin-wave gap at Dirac points, we plot in Fig. \ref{fig2} the constant-energy cuts at different spin-wave energies with an energy integration range of 1.0 meV, obtained by using the $E_i = 25$ meV data. 
Since CrI$_3$ is a ferrromagnet, spin-wave excitations   
stem from the $\Gamma$ point at low energy transfer about $\hbar\omega = 3$ meV 
[Fig. \ref{fig2}(f)]. Upon increasing energy to $\hbar\omega = 6$ meV, spin waves form an isotropic ring pattern around the $\Gamma$ point [Fig. \ref{fig2}(e)]. 
At $\hbar\omega = 10$ meV, the ring breaks into six-folded patterns concentrated around $K/K^\prime$ points,  revealing the typical Heisenberg gap at $M$ [Fig. 2(d)]. When the energy transfer is further increased, the six folded pattern becomes invisible at $\hbar\omega \approx 12$ meV and only reappears for $\hbar\omega \geq 14$ meV.  Consistent with Fig. 1(d), we find that the spin gap around $\hbar\omega \approx 12$ meV extends to the entire BZ including six equivalent Dirac points. 
These results suggest that the associated magnon modes obtain finite mass via interactions with each other or with additional degrees of freedom. One candidate may be the spin-wave interacting with lattice excitations (phonons). 
In general, dynamic spin-lattice coupling can create energy gaps or broadening in the 
magnon dispersion at the nominal intersections of magnon and phonon
modes \cite{anda,Guerreiro,Oh16,Dai,Man}. 
Although CrI$_3$ has several phonon modes in the vicinity of the spin gap \cite{Zhang_CrX3_2015}, 
the large magnitude ($\sim$4 meV) and extension over the entire BZ of the gap suggest that magnon-phonon coupling is unlikely to be the origin of the gap. Alternatively, if one can construct a spin Hamiltonian that breaks the sublattice symmetry of the Cr honeycomb lattice, such a Hamiltonian will feature a magnon spectrum with spin gaps but without nontrivial topology
and chiral edge states expected for topological spin excitations. However, the two Cr sublattice ions are chemically identical in CrI$_3$, and therefore the sublattice
asymmetry spin Hamiltonian can be ruled out for CrI$_3$. 

\begin{figure}[b]
\centering
\includegraphics[width=3.33in]{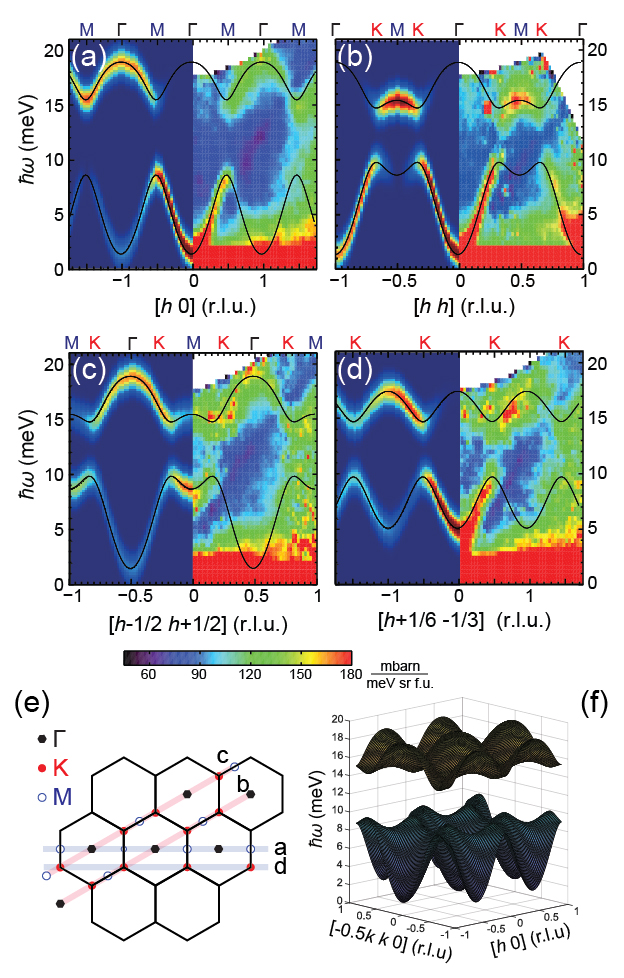}
\caption{\label{fig3} (a-d) The neutron scattering intensities experimentally observed at 5 K are directly compared with the calculated intensities and dispersions discussed in the text. The experimental data on the right panels were integrated over $-5\leq l \leq 5$. The calculations on the left panels used the best fitting parameters with $J_c=0$ as discussed in the text, and convoluted by the effective energy resolution of $\delta(\hbar\omega)$ = 1.8 meV considering the band with along $l$. The displayed cuts are described in (e) within the $(h,k)$ plane. (f) 3D view of the 2D spin-wave excitations of CrI$_3$.}
\end{figure}

To qualitatively understand the observed spin-wave excitations, we fit the spin-wave spectra with the SpinW program \cite{spinw_2015}. By including the DM interaction (${\bf A}$) in the linear spin-wave Heisenberg Hamiltonian as \cite{Owerre_topological_2016}  

\begin{equation}\label{hamiltonian}
H = -\sum_{i<j} \left[ J_{ij} {\bf S}_i\cdot {\bf S}_j + {\bf A}_{ij}\cdot ({\bf S}_i\times{\bf S}_j)\right] - \sum_{j} D_z (S_j^z)^2, 
\end{equation}

\noindent where $J_{ij}$ is magnetic exchange coupling of the 
spin ${\bf S}_i$ and ${\bf S}_j$, ${\bf A}_{ij}$ is the DM interaction between sites $i$ and $j$, and
$D_z$ is the easy-axis anisotropy along the $z$ ($c$) axis. 
We fit the data in the following two steps assuming a Cr spin of $S = 3/2$. First, we integrate the data over $l$ to improve the statistics, and fit the spin-wave dispersions in the $(h,k)$ planes excluding $J_c$.  Assuming that the nearest, next-nearest, and next-next nearest neighbor Cr-Cr magnetic exchange couplings are $J_1$, $J_2$, and $J_3$ as shown in Fig. \ref{fig1}(c), our best-fit values yield $J_1=2.01$, $J_2=0.16$, and $J_3=-0.08$ meV, comparable to the density functional theory calculations \cite{Wang_CrX3_2011,Wang_CrI3_2016,Lado_CrI3_2017}. Remarkably, we find 
$|{\bf A}|$ = 0.31 meV across the next-nearest neighbors, which is larger than $J_2$. We note that the value of $D_z=0.49$ meV obtained by this method is overestimated because there is a finite spin-wave bandwidth along the $l$ direction ($\sim$1.8 meV) as shown in Fig. \ref{fig1}(e). 
Both $J_c$ and $D_z$ were finally obtained by fitting the low energy modes along the $l$ direction while 
fixing the in-plane exchange constants. The best fit values of these two parameters are $J_c=0.59$ and $D_z=0.22$ meV, respectively, which are significantly larger than those in CrBr$_3$ \cite{Samuelsen_CrBr3_1971}. 
In particular, the anisotropy term in CrI$_3$ is an order of magnitude larger than those of CrBr$_3$.
These results suggest that ferromagnetic order in CrI$_3$ has much stronger $c$-axis exchange coupling
compared with CrBr$_3$, although both materials are van der Waals ferromagnets \cite{Liu_CrI3_2018}.  Since the single layer 
CrI$_3$ orders ferromagnetically below about $T_C\approx 45$ K \cite{Huang_CrI3_2017} and not
significantly different from the bulk of $T_C=61$ K, the magnetic ordering
temperature of CrI$_3$ must be mostly controlled by the in-plane magnetic exchange 
couplings as the $c$-axis exchange coupling of $J_c=0.59$ meV in bulk is expected to vanish in the monolayer CrI$_3$.

Since the fitted DM interaction $|{\bf A}|$ and the easy-axis anisotropy $D_z$ are rather large, it is interesting to ask if there are other physical 
effect that may contribute to the observed large spin gap.  
Theoretically, the pseudo-dipolar 
interaction that can arise from super-exchange and spin-orbit
coupling may induce anisotropic bond-directional exchanges
in the nearest neighbor of a honeycomb lattice and open a spin gap 
at the Dirac points \cite{XSWang2017}. However, the effect of pseudo-dipolar interaction on spin gap is much smaller than the comparable strength of DM interaction $|{\bf A}|$. As shown in Fig. 1(b) of Ref. \cite{XSWang2017}, the size of spin gap at Dirac points is rather small even when the pseudo-dipolar interaction is 5 times the nearest neighbor exchange coupling. To induce a spin gap of $\sim$ 4 meV as observed in CrI$_3$, the pseudo-dipolar and single-ion anisotropy terms should be 3.4 and 2.3 times the nearest neighbor exchange coupling, respectively, using the gap formula described in Ref. \cite{XSWang2017}. Therefore, it is highly unlikely that the pseudo-dipolar interaction in CrI$_3$ induces the observed large spin gap at the Dirac points. 

Figure \ref{fig3} compares the calculated spin-wave spectra using these parameters with the experimentally observed dispersions. The left panel in Fig. 3(a) plots the calculated dispersion along the $[h,0]$ direction, while 
the right panel shows the data.  Similar spin-wave calculations and observed spectra 
along the $[h,h]$, $[h-1/2,h+1/2]$, and $[h+1/3,-2/3]$ directions are shown in Figs. 3(b), 3(c), 
and 3(d), respectively. In all cases, the spin-wave gap observed at Dirac $K$ points 
are well reproduced by the calculation.  The calculated spectra 
also reasonably reproduce the strong (weak) intensity of the acoustic (optical) spin-wave modes 
within the 1st BZ, which becomes weaker (stronger) in the 2nd BZ. 
At $M$ points, the ferromagnetic nearest neighbor exchange couplings ($J_1 > 0$) 
of the Heisenberg Hamiltonian 
ensures that the acoustic spin-wave mode is always lower in energy than the optical mode. 
The overall dispersion along the $M$-$\Gamma$-$M$ direction  
is not significantly affected by the DM interaction ${\bf A}$ [Fig. 3(e)]. 
On the other hand, we see clear splitting of the acoustic and optical spin-wave modes
at $K$ points along the $\Gamma$-$K$-$M$ direction due to the large  
DM term, which also enhances the magnon density of states at Dirac points. 
Figure \ref{fig3}(f) shows the calculated overall spin-wave dispersion 
including the DM interaction.

\begin{figure}[b]
\centering
\includegraphics[width=3.33in]{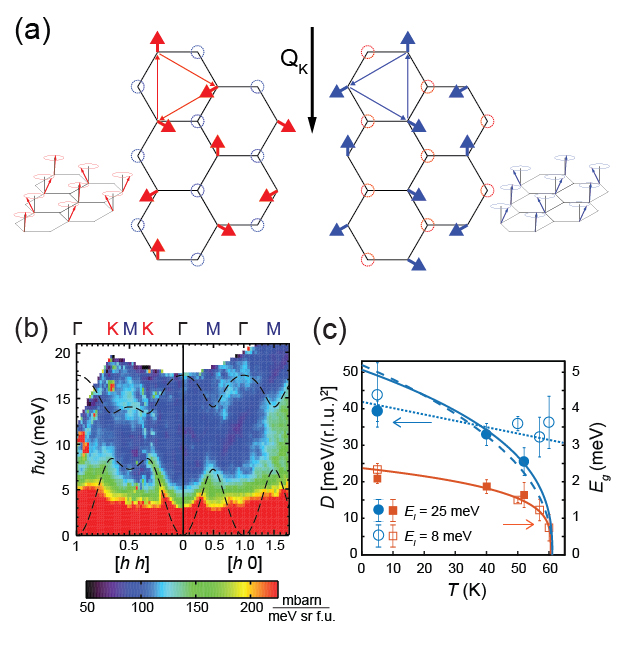}
\caption{\label{fig4} (a) The spin displacements of the two sublattice spins at the right-handed chiral wave at K points. The long arrow at the center denotes the direction of spin-wave propagation. The thin triangular arrows mark the direction of rotation for the right-handed chiral within a hexagon. (b) Spin-wave excitations at $T = 52$ K. The dashed lines are the calculations using the best-fitting parameters from the data at $T$ = 5 K but with $D_z$ set to be 0. (c) Temperature dependencies of the stiffness of the acoustic mode [$D(T)$, circle] and the anisotropy gap ($E_g$, square) obtained by fitting the data integrated over $-5 \le l \le 5$. For this reason, the absolute values of $E_g$ are overestimated by the factor of $\sim$2. The open and closed symbols are the results obtained using $E_i$ = 8 meV and 25 meV, respectively. The blue solid and dashed lines are expected $D(T)$ from 3D Heisenberg and Ising models, respectively.}
\end{figure}

If we assume that the observed spin gap near Dirac points is due to the presence of DM interactions, spin waves in CrI$_3$ should have topological edge states in place of massless Dirac magnons \cite{Chisnell_kagome_2015,Owerre_topological_2016}. Such topological magnons emerge from localizes spin-wave modes forming chiral vortices, 
among which the handedness may be chosen via local magnetic fields. 
From $\Gamma$ towards $K$, the two sublattice spins are displaced from each other along the direction parallel to the wave front. At $K = (1/3, 1/3, 0)$, the spins of one sublattice will precess by 120$^\circ$ along the direction of wave propagation. As a result, the excitations of the two sublattice spins will be decoupled from each other since the in-plane Heisenberg exchanges are frustrated. For instance, the two sublattice excitations illustrated in Figs. \ref{fig4}(a) and \ref{fig4}(b) have the mutually equivalent handedness when they propagate along $K$. When localized within a single hexagon, however, they exhibit opposite handedness from each other as depicted by thin arrows. As a result, these two localized spin-wave excitations can become potentially incompatible with each other. Although mutually degenerate via Heisenberg exchanges, their degeneracy is lifted when the next-nearest neighbor DM vectors are introduced \cite{Owerre_topological_2016}. Interestingly, the apparent magnitude of the DM interactions in CrI$_3$ is larger than $J_2$ along the same next-nearest pair, and as large as 14\% of $J_1$. Given that no such feature has been observed in CrBr$_3$ \cite{Samuelsen_CrBr3_1971}, this effect must arise from 
the larger spin-orbit coupling in CrI$_3$ when Bromine is replaced by the heavier Iodine.

Finally, we discuss temperature dependence of spin-wave excitations in CrI$_3$. As temperature is increased towards $T_C$, the magnons gradually broaden and soften. At $T=52$ K ($T/T_C=0.85$),
the overall spin-wave spectra remain unchanged with a spin gap at Dirac points [Fig. \ref{fig4}(b)].
In the hydrodynamic limit of long wave-lengths and small ${\bf q}$, spin-wave energy 
$\hbar\omega$ has the quadratic $q$ dependence via 
$\hbar\omega=\Delta(T)+D(T)q^2$, where $D(T)$ is temperature dependence of the
spin wave stiffness and $\Delta(T)$ is the small dipolar gap arising
from the spin anisotropy \cite{lovsey}. For a simple 3D Heisenberg ferromagnet, 
temperature dependence of the spin-wave stiffness $D(T)$ is expected to renormalize to zero at $T_C$ via 
$[(T-T_C)/T_C]^{\nu-\beta}$, where $\nu=0.707$ and $\beta=0.367$ are critical exponents \cite{collins}. 
The blue solid and dashed lines in Fig. 4(c) show the resulting temperature dependence of $D(T)$
for 3D Heisenberg and 1D Ising model with $\nu=0.631$ and $\beta=0.326$, respectively.
For CrI$_3$, magnetic critical exponent behavior was found to be 3D-like with
$\beta=0.26$ \cite{Liu_CrI3_2018}.
While temperature dependence of 
$D(T)$ is finite approaching $T_C$ clearly different from the 3D Heisenberg or 1D Ising expectation [Fig. 4(c)], the results are rather similar to $D(T)$ in ferromagnetic manganese oxides
\cite{Dai01}. On the other hand, $\Delta(T)$ vanishes at $T_C$ [Fig. 4(c)], suggesting that the spin anisotropy fields play an important role in stabilizing the 2D ferromagnetic ordering \cite{Lado_CrI3_2017}. 

\section{Conclusions}

In summary, our inelastic neutron scattering experiments  
reveal a large gap in the spin-wave excitations of CrI$_3$ at the Dirac points. 
The acoustic and optical spin-wave bands are separated from each other by $\sim$4 meV, most likely
arising from the next-nearest neighbor DM interaction that breaks  
inversion symmetry of the lattice.  This may lead to a nontrivial topological magnon insulator with magnon edge states, analogous to topological insulators in electronic systems but without electric Ohmic heating. These properties make CrI$_3$ appealing for high-efficiency and dissipationless spintronic applications, 
although there are also challenges in making real spintronic devices  \cite{Chumak,Chernyshev2016,XSWang2017,XSWang2018}. 
Our analysis of the observed spin waves suggests that the DM interaction is stronger than the Heisenberg exchange coupling between the next-nearest neighbor pairs 
in the 2D honeycomb lattice of CrI$_3$. The observation of a large spin gap in magnons of 
CrI$_3$ and its absence in CrBr$_3$ suggests that spin-orbit coupling plays an important
role in the physics of topological spin excitations in honeycomb ferromagnet CrI$_3$.

\section{acknowledgments}
The neutron scattering work at Rice is supported by the US NSF Grant
No. DMR-1700081 (P.D.). The CrI$_3$ single-crystal
synthesis work was supported by the Robert
A. Welch Foundation Grant No. C-1839 (P.D.). The work of J.H.C. was supported by the National Research Foundation of Korea (NRF-2016R1D1A1B03934157; NRF-2017K1A3A7A09016303). Research at Oak Ridge National Laboratory's Spallation Neutron Source was supported by the Scientific User Facilities Division, Office of Basic Energy Sciences, U.S. Department of Energy. Oak Ridge National Laboratory is managed by UT-Battelle, LLC, for U.S. DOE under Contract No. DEAC05-00OR22725.

\end{document}